\documentclass[aps,prl,twocolumn]{revtex4}%
\usepackage{amssymb,amsmath}
\usepackage{bm,url,graphicx,subfigure,epsfig}
\usepackage{color}
\usepackage{natbib}
\usepackage{amsmath}
\usepackage{siunitx}
\usepackage{pxfonts}
\usepackage[T1]{fontenc}
\usepackage{amsfonts}
\usepackage{amssymb}

\newcommand{\beq}{\begin{equation}}
\newcommand{\eeq}{\end{equation}}
\newcommand{\beqa}{\begin{eqnarray}}
\newcommand{\eeqa}{\end{eqnarray}}
\newcommand{\ba}{\begin{array}}
\newcommand{\ea}{\end{array}}

\begin{document}

\title{Bose-Einstein Condensation on the Surface of a Sphere}
\author{A. Tononi}
\affiliation{Dipartimento di Fisica e Astronomia ``Galileo Galilei'',  
Universit\`a di Padova, via Marzolo 8, 35131 Padova, Italy}
\author{L. Salasnich}
\affiliation{Dipartimento di Fisica e Astronomia ``Galileo Galilei'',  
Universit\`a di Padova, via Marzolo 8, 35131 Padova, Italy \\
Istituto Nazionale di Ottica (INO) del Consiglio Nazionale delle 
Ricerche (CNR), \\ via Nello Carrara 1, 50125 Sesto Fiorentino, Italy}

\begin{abstract}
Motivated by the recent achievement of space-based 
Bose-Einstein condensates (BEC) with ultracold alkali-metal atoms under 
microgravity and by the proposal of bubble traps which confine 
atoms on a thin shell, we investigate the BEC thermodynamics 
on the surface of a sphere. We determine 
analytically the critical temperature and the condensate fraction 
of a noninteracting Bose gas. Then 
we consider the inclusion of a zero-range interatomic potential, 
extending the noninteracting results 
at zero and finite temperature. Both in the noninteracting and interacting 
cases the crucial role of the finite radius of the sphere is emphasized, 
showing that in the limit of infinite radius one recovers the familiar 
two-dimensional results. We also investigate the 
Berezinski-Kosterlitz-Thouless transition driven by vortical 
configurations on the surface of the sphere, {analyzing 
the interplay of condensation and superfluidity in this finite-size system.}
\end{abstract}

\maketitle
\noindent {\it Introduction.}
From the theoretical prediction in a series of articles of Bose \cite{bose} and Einstein \cite{einstein1,einstein2}, to its experimental achievement 
in 1995 \cite{cornell,ketterle}, the paradigm of Bose-Einstein 
condensation (BEC) guided and marked the development of a large 
part of modern physics. Despite the fact that a huge variety of phenomena 
emerges from the combination of different trapped configurations 
and a tunable interaction strength \cite{dalfovo}, some basic 
analytical problems of fascinating beauty are currently unexplored.
Among them is the condensation of a bosonic gas of ultracold atoms 
confined on the surface of a sphere. 
Our theoretical study of this system is triggered by the experimental 
possibility to confine the atoms on a spherically symmetric bubble trap 
\cite{garraway}, and by zero-temperature computational works \cite{sun,padavicl,meister} 
on the same topic.
This configuration is produced by a radio frequency dressing 
of the atoms, which allows us to engineer a large variety of radial 
configurations \cite{padavicl,garraway2,garraway3,garraway4,white}.
However, a spherical atom distribution cannot be observed in 
conventional experiments since the atoms fall in the bottom of 
the trap due to gravitational effects \cite{colombe}. Until now, many 
experiments with BEC have been carried on in microgravity settings 
\cite{becker,vanzoest,cho} and bubble trap experiments in microgravity 
are planned for an orbiting cold atom laboratory 
inside the International Space Station \cite{elliott,lundblad,2016APS..DMP.K1118L,lundblad2}.
It is thus pressing to obtain analytical results for these systems, 
whose better understanding would offer an efficient benchmark 
for precise atom interferometry, {improved description of compact stars 
\cite{weber}} and fundamental physics testing \cite{muntinga}.
Here we calculate the critical temperature 
for a BEC of noninteracting bosons confined on the surface of a sphere 
and we derive an expression for their condensate fraction as a 
function of the temperature. Then we consider the 
addition of a zero-range two-body interaction. Within a functional 
integration approach we extend the noninteracting 
results in a Gaussian (one-loop) approximation.
Despite a different topology with respect 
to a planar condensate, it is expected that a thin spherical shell 
undergoes the Berezinski-Kosterilitz-Thouless (BKT) transition \cite{BKT,burt1991}. 
We investigate the relationship between BEC and BKT, whose 
understanding is of general interest for any superfluid system 
on a curved surface described by an angle-valued field \cite{vitelli}. 

\begin{figure}[hbtp] 
\centering
\includegraphics[scale=0.375]{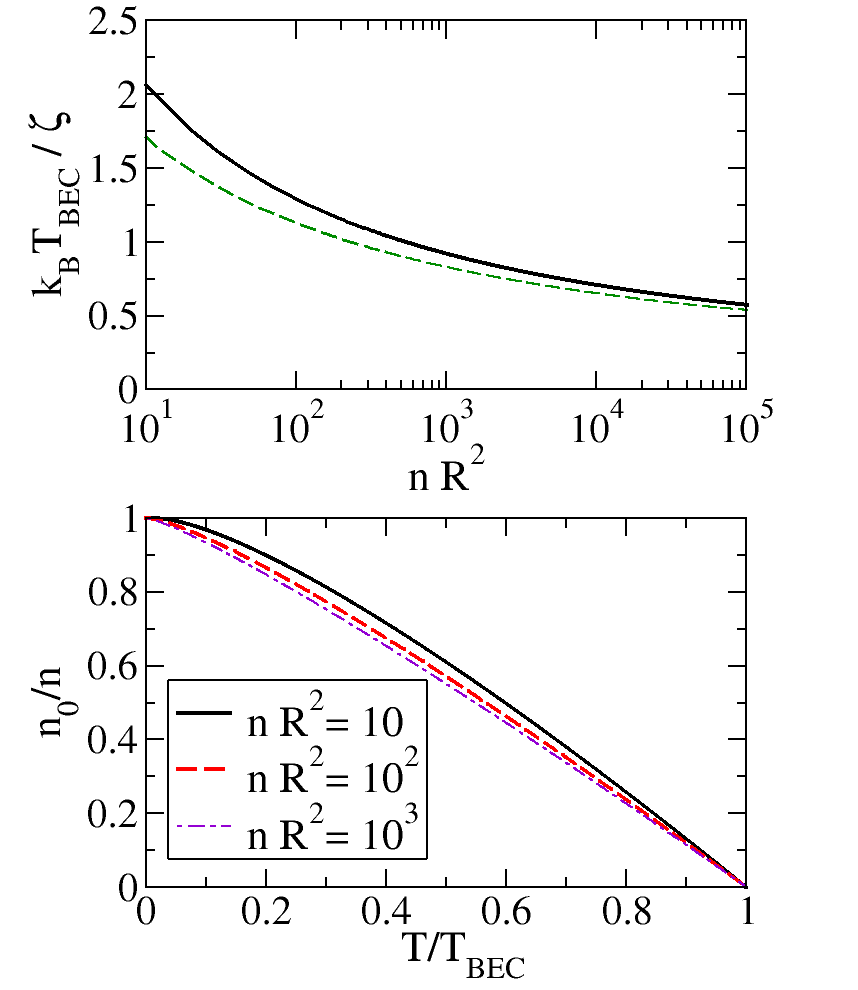} 
\caption{Top: Critical temperature for Bose-Einstein condensation 
in terms of the product $n R^2$, where $R$ is the radius of the sphere
and $k_B T_{\text{BEC}}$ is rescaled with the energy $\zeta = \hbar^2 n/m$.
{Notice how our result in the semiclassical approximation (solid line) 
converges for $nR^2 \gg 1$ to the numerical evaluation of the summation of 
Eq. (\ref{density1}) (dashed line).}
As expected, for fixed $n$ the critical temperature tends to zero in the 
thermodynamic two-dimensional limit $nR^2 \rightarrow \infty$.
Bottom: Condensate fraction $n_0/n$ for noninteracting bosons 
on the sphere surface in terms of the rescaled temperature $T/T_{\text{BEC}}$, 
obtained from the numerical solution of Eq. (\ref{condfracnonint}) with $n R^2$ fixed.}
\label{fig:1}
\end{figure}

\noindent {\it Non-interacting Bose gas.}
The energy of a particle of mass $m$ moving on the surface 
of a sphere of radius $R$ is quantized according to the formula 
\beq 
\epsilon_{l} = {\hbar^2\over 2m R^2} l(l+1) \; , 
\eeq
where $\hbar$ is the reduced Planck constant and $l=0,1,2,...$ is the 
integer quantum number of the angular momentum. This energy level 
has the degeneracy $2l+1$ due to the magnetic quantum 
number $m_l=-l,-l+1,...,l-1,l$ of the third component of the 
angular momentum. In quantum statistical mechanics 
the total number $N$ of noninteracting bosons moving 
on the surface of a sphere and at equilibrium 
with a thermal bath of absolute temperature $T$ is given by 
\beq 
N = \sum_{l=0}^{+\infty} {2l+1\over e^{(\epsilon_l -\mu)/(k_B T)}-1} \; , 
\eeq
where $k_B$ is the Boltzmann constant and $\mu$ is the chemical potential. 
In the Bose-condensed phase, we can set $\mu=0$ and 
\beq 
N = N_0 + \sum_{l=1}^{+\infty} {2l+1\over e^{\epsilon_l/(k_B T)}-1} \; , 
\label{density1}
\eeq
where $N_0$ is the number of bosons in the lowest 
single-particle energy state, i.e. the number of bosons in the Bose-Einstein 
condensate (BEC). From this equation one gets a critical temperature $T_{\text{BEC}}$ 
above which $N_0=0$. Strictly speaking, in our system with a finite 
radius $R$ and finite particle number $N$, $\mu$ cannot be zero and 
there is never a full depletion of the condensate above the critical temperature. 
This residual population of the condensate is however rapidly vanishing 
for a finite but macroscopic system.
Within the semiclassical approximation, 
where $\sum_{l=1}^{+\infty} \to \int_1^{+\infty} dl$, 
Eq. (\ref{density1}) becomes 
\beq 
n = n_0 + {m k_B T\over 2\pi\hbar^2} \left( 
{\hbar^2\over mR^2 k_BT} - 
\ln{\left( e^{\hbar^2/(mR^2 k_B T)} -1 \right)} \right) \, , 
\label{equat1}
\eeq
where $n=N/(4\pi R^2)$ is the 2D number density and $n_0=N_0/(4\pi R^2)$ 
is the 2D condensate density. 
We emphasize that in the low-temperature limit of $T \rightarrow 0$ the 
second term of Eq. (\ref{equat1}) vanishes and the system density $n$ is 
coincident with the condensate density $n_0$.
At the critical temperature $T_{\text{BEC}}$ the condensate density must be zero: 
from Eq. (\ref{equat1}) one finds 
\beq 
k_B T_{\text{BEC}} = { {2\pi \hbar^2\over m} n \over 
{\hbar^2\over mR^2 k_B T_{\text{BEC}}} - \ln{( 
e^{\hbar^2/(mR^2 k_B T_{\text{BEC}})} -1 )} } \; , 
\label{equat2}
\eeq
which is an implicit analytical formula for the critical temperature 
$T_{\text{BEC}}$ as a function of the 2D number density $n$ and the radius $R$ of 
the sphere. As expected \cite{mermin}, in the limit $R\to +\infty$ one gets $T_{\text{BEC}}\to 0$. 
However, for any finite value of $R$ the critical 
temperature $T_{\text{BEC}}$ is larger than zero. This can also be seen in the 
top panel of Fig. \ref{fig:1}, where we report the critical temperature 
$T_{\text{BEC}}$ for noninteracting bosons as a function of the parameter $n R^2$.
The semiclassical approximation (solid line) works very 
well because the strong convergence to zero of the Bose distribution for 
high values of $l$ cuts off the pathological behavior of the density of 
states for $l \gg 1$ \cite{bagnato}. 
Combining Eqs. (\ref{equat1}) and (\ref{equat2}) one immediately 
obtains the condensate fraction of the system for $0\leq T \leq T_{\text{BEC}}$, namely 
\beq \label{condfracnonint}
{n_0\over n} = 1 - { {1 - k_B T {mR^2\over \hbar^2} 
\ln{\left( e^{\hbar^2/(mR^2 k_BT)} -1 \right)}} 
\over 
{1 - k_B T_{\text{BEC}} {mR^2\over \hbar^2} 
\ln{\left( e^{\hbar^2/(mR^2 k_B T_{\text{BEC}})} -1 \right)}} } \; . 
\eeq
The numerical solution of Eq. (\ref{condfracnonint}) 
is reported in the bottom panel of Fig. \ref{fig:1}, in which we 
represent the condensate fraction $n_0/n$ of noninteracting bosons 
in terms of the rescaled temperature $T/T_{\text{BEC}}$ for different values 
of the $n R^2$ parameter, with $n=N/(4 \pi R^2)$.
Experimentally, one can tune $n R^2$ simply changing the total number $N$, 
but also changing the radius $R$ at fixed density $n$.

\noindent {\it Interacting Bose gas. }
We now consider a system of interacting bosons on the surface of a sphere.
The main thermodynamic function describing a system of  
particles in the grand canonical ensemble is the grand potential 
$\Omega= - \beta^{-1} \ln({ \cal Z })$,
where $\beta^{-1}=k_{B}T$ and ${\cal Z}$ is the grand canonical 
partition function. Within the formalism of functional integration 
we calculate ${\cal Z}$ as the functional integral, 
\beq
\label{partfunction}
{ \cal Z }=\int{{ \cal D} [\bar{\psi},\psi] \; 
e^{-S[\bar{\psi},\psi]/ \hbar}},
\eeq
where 
\beq
\label{action}
S[\bar{\psi},\psi] = \int_{0}^{\beta\hbar} d\tau \, 
\int_{0}^{2 \pi} d \varphi \, \int_{0}^{\pi} \sin(\theta) 
\; d \theta \, R^2 \,{ \cal L }(\bar{\psi},\psi) 
\eeq
is the Euclidean action, and
\beq
\label{lagr} 
{ \cal L } = \bar{\psi}(\theta,\varphi,\tau) 
\bigg(\hbar\partial_{\tau}+\frac{L^2}{2mR^2}-\mu\bigg) 
\psi(\theta,\varphi,\tau) + \frac{g}{2} |\psi(\theta,\varphi,\tau)|^{4}
\eeq 
is the Euclidean Lagrangian, i.e. the Lagrangian with imaginary time $\tau$.
Notice that the kinetic energy is written in terms of the angular momentum 
$L$, which is proportional to the angular components of the Laplacian 
operator in spherical coordinates, namely, 
\beq
L^2 = -\hbar^{2} \bigg( \frac{1}{ \sin(\theta)} 
\frac{\partial^{2}}{\partial \varphi^{2}} + 
\frac{1}{\sin^{2}(\theta)} \frac{\partial}{\partial \theta} 
\bigg( \sin(\theta) \, \frac{\partial}{\partial \theta} \bigg) \bigg).
\eeq
Our Lagrangian models interacting bosons on the surface of a sphere.
Experimentally, this corresponds to the thin-shell limit of a bubble 
trap potential \cite{garraway3}, in which the atoms are confined by 
the radial shifted harmonic potential $V_{shell} = m \omega_{\text{sh}}^2 (r-R)^2/2$. 
Following Refs. \cite{garraway2,sun}, we suggest that one can tune the experimental parameters 
to confine the atoms on a shell with a radial harmonic length $l_{\text{sh}}=(\hbar/(m \omega_{\text{sh}}))^{1/2}$ of the order 
$l_{\text{sh}} \approx 0.1 \, \mu \text{m}$, which is much smaller than the radius of the sphere $R \approx 10 \, \mu \text{m}$. 
In this case the radial excitations are inhibited and 
it is safe to impose our constraint in the Lagrangian of Eq. (\ref{lagr}).
Therein, the real two-body interatomic potential $V(r)$ has been 
substituted with the effective two-dimensional zero-range interaction coupling $g$.
Therefore, for a given interatomic potential, one can calculate the 
exact value of $g$ \cite{nota}. One can also use the scattering theory to link 
$g$ to the two-dimensional s-wave scattering length $a_s$. In $\text{2D}$ the latter quantity 
is however energy-dependent and a physical cutoff, usually identified with the 
chemical potential of the system, is needed \cite{astrakharchik}. 
Besides this, the relation between $g$ and $a_s$ should also include the 
corrections due to the curvature of the scattering surface \cite{zhang2018}.
To keep the compatibility with different experimental 
setups and interparticle interactions, in the following we will simply employ $g$,
which could also be used as a phenomenological fitting parameter for 
thicker spherical shells.

Let us now explicitly perform the functional integration of the 
Lagrangian ${ \cal L }$.
The spontaneous breaking in the condensate phase of the $\text{U}(1)$ 
symmetry of the complex order parameter $\psi$ is introduced 
with the Bogoliubov shift
\beq
\label{parametrizationveta}
\psi(\theta,\varphi,\tau) = \psi_{0} + \eta(\theta,\varphi,\tau),
\eeq
where the real field configuration $\psi_{0}$ describes the condensate 
component with angular momentum $l=0$ and $m_{l}=0$.
By substituting this field parametrization and keeping only second 
order terms in the field $\eta$ we rewrite the Lagrangian as 
\beq
{ \cal L } = { \cal L }_{0} + { \cal L }_{g},
\eeq
with ${ \cal L }_{0} = -\mu \psi_{0}^2 +g \psi_{0}^4 /2$, and 
\beqa \label{gausslagr}
{ \cal L }_{g} = \bar{\eta}(\theta,\varphi,\tau) \, \bigg(\hbar \partial_{\tau}  \nonumber
+\frac{L^2}{2mR^2}-\mu\bigg) \, \eta(\theta,\varphi,\tau) + \\ \frac{g}{2} 
[\bar{\eta}^{2}(\theta,\varphi,\tau) + \eta^{2}(\theta,\varphi,\tau)].
\eeqa
The mean-field Lagrangian ${ \cal L }_{0}$ gives the mean-field grand 
potential
\beq \label{mfgp}
\Omega_{0}= 4 \pi R^2 \big( -\mu \psi_{0}^2 +g \psi_{0}^4 /2 \big), 
\eeq
while the functional integral of the Gaussian Lagrangian ${ \cal L }_{g}$ 
can be calculated explicitly with the following decomposition of 
the complex fluctuation field $\eta(\theta,\varphi,\tau)$, 
\begin{eqnarray}
\eta(\theta,\varphi,\tau)= \sum_{\omega_{n}} \sum_{l=1}^{\infty} 
\sum_{m_{l}=-l}^{l} \frac{e^{-i\omega_{n}\tau}}{R} {\cal Y}_{m_{l}}^{l}(\theta,\varphi) 
\, \eta(l,m_{l},\omega_{n}), 
\end{eqnarray}
and similarly for $\bar{\eta}(\theta,\varphi,\tau)$, where $\omega_n$ 
are the Matsubara frequencies, and we introduce the orthonormal basis 
of the spherical harmonics ${\cal Y}_{m_{l}}^{l}$ \cite{brink}.
Substituting this decomposition into the Gaussian Lagrangian (\ref{gausslagr}) 
and using the orthonormality properties of ${\cal Y}_{m_{l}}^{l}$ and 
of the complex exponentials, we rewrite the Gaussian action $S_{g}$ as
\beq
S_{g}[\bar{\eta},\eta]=\frac{\hbar}{2} \sum_{\omega_{n}} \sum_{l=1}^{\infty} 
\sum_{m_{l}=-l}^{l}
\begin{pmatrix}
\bar{\eta}(l,m_{l},\omega_{n}) \\ \eta(l,-m_{l},-\omega_{n})
\end{pmatrix}^{T}\textbf{M}
\begin{pmatrix}
\eta(l,m_{l},\omega_{n}) \\
\bar{\eta}(l,-m_{l},-\omega_{n})
\end{pmatrix},
\eeq
where the elements $\textbf{M}_{ij}$ of the matrix $\textbf{M}$ are defined as
\begin{align}
\begin{split}
\textbf{M}_{ii} \; &= (-1)^{i} i\hbar\omega_{n} + \epsilon_l 
- \mu + 2 g \psi_{0}^2, \quad i=1,2 \\
\textbf{M}_{12} &= \, \textbf{M}_{21} = (-1)^{m} g \psi_{0}^2.
\end{split}
\end{align}
The Gaussian action $S_{g}$ can be integrated in the the 
$\omega_{n} \ge 0$ field sector and the corresponding contribution 
to the Gaussian grand potential $\Omega_{g}$ reads
\beq
\label{omegag}
\Omega_{g}(\mu,\psi_0^2) = \frac{1}{2 \beta} \sum_{\omega_{n}} 
\sum_{l=1}^{\infty} 
\sum_{m_{l}=-l}^{l} \ln \{\beta^2[\hbar^2\omega_{n}^2+E_{l}^2(\mu,\psi_0^2)] \},
\eeq
where $E_{l}(\mu,\psi_0^2)$ is the excitation spectrum of the interacting system:  
\beq
E_{l}(\mu,\psi_0^2)=\sqrt[]{(\epsilon_l
-\mu + 2 g \psi_{0}^2 )^{2}-g^2 \psi_{0}^4}.
\eeq
One can easily sum over the Matsubara bosonic frequencies $\omega_n$ \cite{tononi1}, 
and remembering the mean-field grand potential $\Omega_{0}$ of 
Eq. (\ref{mfgp}), we obtain the total grand potential $\Omega$ as
\beqa
\label{grandpotential2}
\Omega(\mu,\psi_0^2) &=& 4 \pi R^2 \big( -\mu \psi_{0}^2 +g \psi_{0}^4 /2 \big) 
+  \frac{\alpha}{2} \sum_{l=1}^{\infty} \sum_{m_{l}=-l}^{l} \, 
E_{l}(\mu,\psi_0^2)\nonumber \\ &+& \frac{\alpha}{\beta}  \sum_{l=1}^{\infty} 
\sum_{m_{l}=-l}^{l} \, \ln(1-e^{-\beta E_{l}(\mu,\psi_0^2)}) + o(\alpha^2),
\eeqa
where we include the parameter $\alpha=1$, whose power counts the perturbative 
order of the grand potential expansion \cite{kleinert2004,kleinert2005}.
We fix the value of the order parameter $\psi_0$ with the variational 
saddle-point 
condition $\partial \Omega/\partial \psi_0  = 0$, which 
determines a relation between $\psi_0$, the chemical potential $\mu$ 
and the contact interaction strength $g$. 
Since the condensate density $n_0$ is defined as $n_0=\psi_0^2$, 
the saddle-point condition implies that 
\beq \label{condensatedensity}
n_0(\mu) = \frac{\mu}{g} - \frac{\alpha}{4 \pi R^2} \sum_{l=1}^{\infty} 
\sum_{m_{l}=-l}^{l} \, \frac{\frac{\hbar^2 l (l+1)}{2 m R^2} + \mu}{E_{l}^{B}} 
\, \bigg[\frac{1}{2} + \frac{1}{e^{\beta E_{l}^{B} }-1}  \bigg]+ o(\alpha^2),
\eeq
which at the lowest perturbative order gives $n_0 = \mu/g + o(\alpha)$, 
and where the excitation spectrum $E_{l}^{B}=E_{l}[\mu,n_0(\mu)]$ 
takes the Bogoliubov-like form \cite{bogoliubov1947}
\beq
\label{excitationspectrum}
E_{l}^{B}=\sqrt[]{\epsilon_l 
(\epsilon_l +  2 \mu ) }.
\eeq
With the mean-field condition, the grand potential of Eq.
(\ref{grandpotential2}) can be rewritten as
\beqa
\label{grandpotential3}
\Omega[\mu,n_0(\mu)] 
&=& -4 \pi R^2 \frac{\mu^2}{2 g} + \frac{\alpha}{2} \sum_{l=1}^{\infty} 
\sum_{m_{l}=-l}^{l} \, E_{l}^{B} \nonumber \\ &+& \frac{\alpha}{\beta}  
\sum_{l=1}^{\infty} \sum_{m_{l}=-l}^{l} \, \ln(1-e^{-\beta E_{l}^{B}}) 
+ o(\alpha^2),
\eeqa
and we can introduce the system density $n$ as 
\beq
n(\mu) = - \frac{1}{4 \pi R^2} \frac{\partial \Omega(\mu,n_0(\mu))}
{\partial \mu}.
\eeq
Since we are interested in the relation between the density $n$ and 
the condensate density $n_0$, we substitute in the last equation the 
value of $\mu=\mu(n_0)$ given by Eq. (\ref{condensatedensity}), obtaining
\beq \label{densityn0}
n(n_{0})= n_{0} + f_{g}^{(0)}(n_0) + f_{g}^{(T)}(n_0),
\eeq
where
\beq \label{fg0}
f_{g}^{(0)}(n_0) = \frac{\alpha}{4 \pi R^2} \frac{1}{2} \sum_{l=1}^{\infty} 
\sum_{m_{l}=-l}^{l} \, \frac{\epsilon_l + g n_0}{ E_{l}[ \mu(n_0),n_0 ]} 
\eeq
is the zero-temperature Gaussian density and
\beq \label{fgT}
f_{g}^{(T)}(n_0) = \frac{\alpha}{4 \pi R^2} \sum_{l=1}^{\infty} 
\sum_{m_{l}=-l}^{l}  \, \frac{\epsilon_l + g n_0}{ E_{l}[ \mu(n_0),n_0 ]} 
\, \frac{1}{e^{\beta E_{l}[ \mu(n_0),n_0 ]}-1}
\eeq
is the finite-temperature Gaussian density.
We emphasize that, at a Gaussian level, this procedure is equivalent 
to the one adopted in our previous article \cite{tononi2018}.
The number density of Eq. (\ref{densityn0}) constitutes a reliable 
result in a low-temperature and weakly interacting regime, in which 
the quantum and thermal depletion are small, namely $n \approx n_0$.
Within the variational perturbation theory (VPT) \cite{kleinertbook,kleinertbook2}, 
one can employ the weakly interacting perturbative expansion of 
Eq. (\ref{densityn0}) to derive a self-consistent approximation 
for the number density, valid also for larger values of the depletion.
In our case, the VPT procedure outlined in Ref. \cite{kleinert2005} is 
equivalent at the order $o(\alpha^2)$ to the substitution of the 
condensate density $n_{0}$ in Eqs. (\ref{fg0},\ref{fgT}) 
with the total number density $n${, obtaining $f_{g}^{(0)}(n)$
 and $f_{g}^{(T)}(n)$. We stress that this 
VPT method for a 3D homogeneous Bose gas leads to a critical 
temperature that scales with the square root of the gas parameter, 
while Monte Carlo simulations suggest a linear scaling \cite{gruter}.} 
Setting $\alpha=1$, we now calculate explicitly $f_{g}^{(0)}(n)$,
which is ultraviolet divergent and needs a regularization procedure. 
We rewrite the sum as an integral over $l$ 
in which the degeneration over $m_l$ gives a $2l+1$ factor.
Using the variable $t=\hbar^2 l (l+1)/(4 m n g R^2)$, we integrate 
$f_{g}^{(0)}(n)$ subtracting the pathological asymptotic behavior 
of the integrand function at $+\infty$, thus obtaining
\beq \label{fg0final}
f_{g}^{(0)}(n) = \frac{m g n}{4 \pi \hbar^2} + \frac{1}{4 \pi R^2} \bigg[ 
1 - \sqrt{1+\frac{2 g m n R^2}{\hbar^2}} \bigg],
\eeq
which vanishes for noninteracting bosons for which the quantum 
depletion does not occur.
We emphasize that $f_{g}^{(0)}(n)$ generalizes the quantum depletion 
result by Schick for a weakly interacting Bose gas in $\text{2D}$ \cite{schick1971}, 
by including a correction due to the finite size of the sphere radius.
In particular, Schick's result is reproduced in the 
$R \rightarrow \infty$ limit in which the interaction coupling 
$g$ can be identified with $g=2 \pi \hbar^2/[m |\ln(n a_s^2)|]$,
where $a_s$ is the two-dimensional s-wave scattering length \cite{astrakharchik}.
Similarly, we calculate the thermal density $f_{g}^{(T)}(n)$, obtaining
\beq \label{fgTfinal}
f_{g}^{(T)}(n) =\frac{1}{2 \pi R^2} 
\sqrt{1+\frac{2 g m n R^2}{\hbar^2}} - \frac{m k_{B} T}{2 \pi \hbar^2} 
\ln \bigg( e^{\frac{\hbar^2}{m R^2 k_{B} T} \sqrt{1+(2 g m n R^2/\hbar^2)}}-1\bigg).
\eeq
Putting together the density contributions Eqs. (\ref{fg0final}) and (\ref{fgTfinal}) 
with $n_0$, we get the VPT-improved self-consistent 
condensate fraction of an interacting Bose gas on the surface of a sphere as
\beqa \label{finaldensityVPT}
{n_0\over n} &=& 1 - \frac{m g}{4 \pi \hbar^2} - \frac{1}{4 \pi R^2 n} 
\bigg[ 1 + \sqrt{1+\frac{2 g m n R^2}{\hbar^2}} \bigg] \nonumber 
\\ &+& \frac{m k_{B} T} {2 \pi \hbar^2n} 
\ln \bigg( e^{\frac{\hbar^2}{m R^2 k_{B} T} 
\sqrt{1+(2 g m n R^2/\hbar^2)}}-1\bigg).
\eeqa
With $n_0=0$ in Eq. (\ref{finaldensityVPT}), we calculate an implicit relation 
for the condensation critical temperature of interacting bosons 
[with $\beta_{BEC}=(k_{B}T_{\text{BEC}})^{-1}$]
\beq \label{Tcinteracting}
k_B T_{\text{BEC}} = \frac{\frac{2 \pi \hbar^2 n}{m} - \frac{g n}{2} }{ 
\frac{\hbar^2 \beta_{BEC}}{2 m R^2} \bigg( 1 + \sqrt{1+\frac{2 g m n R^2}{\hbar^2}} \bigg) 
- \ln \bigg( e^{\frac{\hbar^2\beta_{BEC}}{m R^2} 
\sqrt{1+\frac{2 g m n R^2}{\hbar^2}}}-1\bigg)}.
\eeq
Note that the critical temperature for the noninteracting system 
of Eq. (\ref{equat2}) is reproduced if $g=0$ is set.
\begin{figure}[hbtp]
\centering 
\includegraphics[scale=0.375]{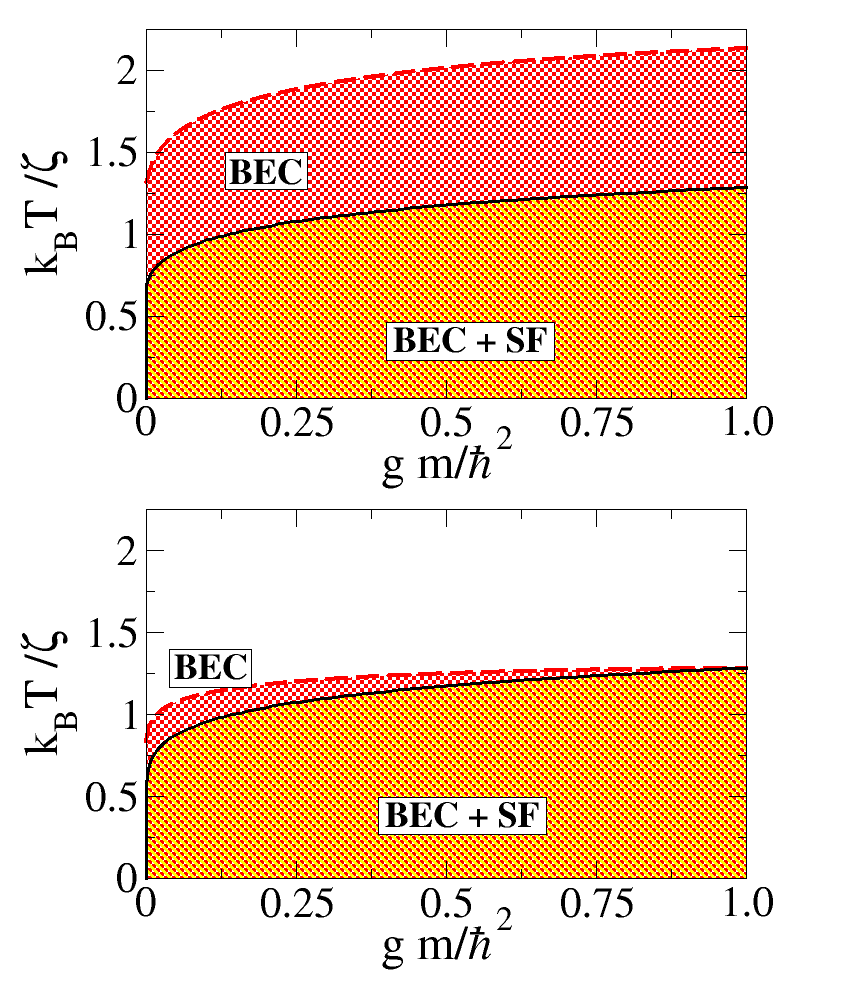}
\caption{
Phase diagram of the system for two values of $n R^2$:
$10^2$ in the upper panel, $10^4$ in the lower panel.
The dashed lines represent the critical temperature $k_B T_{\text{BEC}}/\zeta$, 
rescaled with the typical energy $\zeta= \hbar^2 n /m$, plotted in terms 
of the adimensionalized zero-range interaction strength $g m/\hbar^2$.
The solid curve represents the critical temperature $k_B T_{BKT}/\zeta$ 
of the Berezinski-Kosterlitz-Thouless transition. 
{Notice that, depending on the values of $nR^2$ and of 
$g m/\hbar^2$ and within the approximations adopted (see text), the system can show 
coexistence of condensation and superfluidity (BEC+SF), or condensation 
in the absence of superfluidity (BEC).}
} \label{fig:2}
\end{figure}
In Fig. \ref{fig:2} we report the critical temperature $k_B T_{\text{BEC}}/\zeta$ (dashed line), 
rescaled with the energy $\zeta=\hbar^2 n /m$, in terms of $g m/\hbar^2$.
The shaded area under each of the dashed curves with 
$n R^2$ fixed is where BEC occurs:
if the density $n$ is kept fixed and the sphere radius $R$ 
is increased, this area diminishes.
Indeed, the expected result of $T_{\text{BEC}} = 0$ is reobtained in the 2D flat-system 
limit $R \rightarrow \infty$. 

A spherical surface is topologically inequivalent to the 2D flat plane. 
In particular, the presence of a point at the infinity allows 
only for couples of topological defects to exist, being 
vortex-antivortex dipoles, or free vortices {\cite{hairyball}}. 
Despite this fact, the Kosterlitz-Nelson criterion \cite{nelson} 
for the jump of the superfluid density $n_{s}(T)$ was recovered extending the 
Berezinski-Kosterlitz-Thouless theory on the sphere \cite{burt1991}: 
i.e., $k_{B}T_{BKT}/(\hbar^2 n_{s}(T_{BKT})/m)=\pi /2$. 
{Here, in analogy to the Landau formula for the 2D plane, we calculate $n_s(T)$ as 
\beq
n_s= n-\frac{1}{k_{B}T} \int_{1}^{+\infty} \frac{d l \, (2l+1)}{4 \pi R^2} \, 
\frac{\hbar^2 (l^2+l)}{2 m R^2} \frac{e^{E_l^B /(k_{B}T)}}{(e^{ E_l^B /(k_{B}T)}-1)^2}, 
\label{ns}
\eeq }
and applying the Kosterlitz-Nelson criterion we evaluate numerically 
the critical temperature $T_{BKT}$, 
{represented as the solid curve of Fig. \ref{fig:2}. 
We find that $T_{BKT}$ has a weak dependence on $nR^2$ and goes to 
zero in an exponentially small region where $gm/\hbar^2 \to 0$: 
this result is in agreement with the classical field simulations 
of Ref. \cite{prokofev} for bosons in a 2D uniform configuration. 
Indeed, the spherical surface is locally isomorphic to the Euclidean plane \cite{burt1991}, 
where superfluidity is absent in the noninteracting limit. 
At the same time, we stress that for any tiny but physically meaningful 
interaction strength, the critical BKT temperature $T_{BKT}$ is practically 
finite, while $T_{\text{BEC}}$ coincides with the noninteracting one. 
In this regime of vanishing interaction, the unification of BEC and BKT 
transitions observed with bosons in a 2D harmonic trap \cite{fletcher}, 
is obtained only when $nR^2 \gtrsim 10^4$.}

{ 
In the phase diagram of Fig. \ref{fig:2}, within the approximations involved 
in the calculation of $n_s(T)$ and at the zero order of VPT, we find a region 
where condensation and superfluidity coexist, and a region where condensation 
is not accompanied by superfluidity. 
Note that the latter condition is more pronounced for $nR^2 \lesssim 10^2$ 
and was experimentally observed in a quasi-2D finite-size Bose gas \cite{clade}: 
at low density, the curvature of the sphere may play the same role of their 
2D weak external potential. 
Finally, in the regime of $gm/\hbar^2 \gg 1$ where our perturbative scheme 
is not expected to hold, we point out that $T_{BKT}$ becomes essentially constant 
while $T_{\text{BEC}} \to 0$ due to a large depletion of the condensate. 
The liability of this result should be established with more refined methods. 
}

\noindent {\it Conclusions.}
The condensate fraction for noninteracting and interacting bosons 
on the surface of a sphere, i.e. Eqs. (\ref{equat2}), (\ref{condfracnonint}), 
(\ref{finaldensityVPT}) and (\ref{Tcinteracting}), can be experimentally observed 
in microgravity conditions with bubble traps in the thin-shell limit. 
{These results are concrete predictions to test quantum statistical 
mechanics in regimes where finite-size and curvature effects play a relevant role. 
Further experimental and theoretical investigations should also focus on
the expected interplay of superfluidity and Bose-Einstein condensation.
We expect that a typical 2D configuration, with $N \approx 10^5$ $^{87}$Rb atoms confined 
on a shell with radius $R = 10 \, \mu \mbox{m}$ and thickness $l_{\text{sh}} = 0.1 \, \mu\mbox{m}$, 
has a critical temperature of $T_{\text{BEC}} = 670 \, \text{nK}$ for 
$g m/\hbar^2 = 2^{3/2}\pi^{1/2}a_{s}/l_{\text{sh}} = 0.26$, where $a_{s}=5 \, \text{nm}$ 
is the bare s-wave scattering length. 
With the Feshbach resonance one can tune $a_{s}$ and investigate also 
regimes with higher values of $gm/\hbar^2$.}

\begin{acknowledgements}
The authors acknowledge {G. Bighin}, {F. Cinti}, 
{B. Garraway}, {T. Macri}, D. Partipilo, 
{A. Pelster}, and F. Toigo for useful suggestions and discussions.
\end{acknowledgements}


\begin{thebibliography}{99}
\bibitem{nota} Our two-dimensional strength $g$ is defined as 
$\int d^{3} r V(r) \delta(r-R)$, where $V(r)$ is the three-dimensional 
two-body interaction. Given a Gaussian interaction potential 
$V(r)=V_{0} \, e^{-r^2/\sigma^2}$, we find 
$g=V_{0} \pi \sigma^2 (1 - e^{-4R^2/\sigma^2})$.
For the Yukawa potential $V(r)= -V_{0}e^{-r/\lambda}/r$ one finds 
$g=-2 \pi V_0 \lambda R (1- e^{-2R/\lambda})$, while 
a short-range box potential $V(r)=-V_{0} \, \Theta(b-r)$ gives  
$g=-2 \pi V_0 b^2$.
\end{thebibliography}
\end{document}